\documentclass[showpacs,aps,graphicx]{revtex4}
\usepackage{graphicx}

\begin{document}

\title{Optimal concentrating arbitrary partially entangled W states with linear optics}

\author{Lan Zhou$^{1,2}$, Yu-Bo Sheng$^{2,3}$\footnote{Email address:
shengyb@njupt.edu.cn} }
\address{$^1$ College of Mathematics \& Physics, Nanjing University of Posts and Telecommunications, Nanjing,
210003, China\\
$^2$Key Lab of Broadband Wireless Communication and Sensor Network
 Technology, Nanjing University of Posts and Telecommunications, Ministry of
 Education, Nanjing, 210003, China\\
$^3$  Institute of Signal Processing  Transmission, Nanjing
University of Posts and Telecommunications, Nanjing, 210003,  China\\}

\date{\today}

\begin{abstract}
We proposed two optimal entanglement concentration protocols (ECPs) for arbitrary single-photon multi-mode W state and multi-photon polarization W state, respectively. In both ECPs, we only require one pair of partially entangled W state, and do not consume any auxiliary photon. Both ECPs are based on the linear optics which can be easily realized.
On the other hand, the concentrated maximally entangled states can be remained, which are quite different from the previous ECPs.
Moreover, for the concentration of the arbitrary single-photon N-mode W state or N-mode polarization W state, the total success probability
is equal to  Nth the modulus square of the Schmidt coefficient of the smallest magnitude. It makes both ECPs optimal than all the previous ECPs.
Our ECPs may be useful in current quantum communication fields.

\end{abstract}
 \maketitle

\section{Introduction}

   In the recent years, quantum information processes have developed rapidly \cite{book}, among which the most important branches
are the long-distance quantum communication and quantum computation. Entanglement, which is a uniquely quantum mechanical
feature, is considered to be an essential resource for both the two branches. In practical applications, entanglement is usually produced locally and can be distributed to the remote parties. It not only can hold the power for the quantum nonlocality \cite{Einstein}, but also can provide wide applications in the quantum information processing (QIP) \cite{rmp}. For example, many popular research areas such as the quantum teleportation \cite{teleportation,cteleportation1,cteleportation2}, quantum denescoding \cite{densecoding},  quantum secret sharing\cite{QSS1,QSS2,QSS3}, quantum state
sharing\cite{QSTS1,QSTS2,QSTS3}, and quantum secure direct communication \cite{QSDC1,QSDC2,QSDC3}, all require entanglement to set up the quantum entanglement channels.

Among various entanglement forms, the multi-mode and multi-particle W states have quite important applications. The perfect entangled W states are the maximally entangled W state, which can be written as
\begin{eqnarray}
|W\rangle_{multi-mode}&=&\frac{1}{\sqrt{N}}(|100\cdots 0\rangle+|010\cdots 0\rangle+|001\cdots 0\rangle+\cdots+|000\cdots 01\rangle),\nonumber\\
|W\rangle_{muti-photon}&=&\frac{1}{\sqrt{N}}(|HVV\cdots V\rangle+|VHV\cdots V\rangle+|VVH\cdots V\rangle+\cdots+|VVV\cdots VH\rangle),
\end{eqnarray}
 where the $|H\rangle$ and $|V\rangle$ represent the horizontal and vertical polarization of the photon state, while $|1\rangle$ and $|0\rangle$ represent one photon and no photon, respectively. It has been proved that the W states are highly robust against the loss of one or two qubits \cite{W,W1,W2}. There are many works have been done based on both multi-particle W state and single-photon multi-mode W state, such as the protocols of perfect teleportation and superdense coding with W states \cite{wteleportation},
  the generation of the W state \cite{generation1,generation2,generation3,generation4,generation5,generation6,generation7,generation8,generation9,generation10,generation11}, entanglement transformation \cite{wtransformations},
  distillation \cite{wdistill1,wdistill2,wdistill3,cao} and concentration \cite{zhanglihua,wanghf,shengPRA,gub,duff,zhoujosa,zhouqip2} of the W states.  Interestingly, Gottesman \emph{et al.} proposed a protocol for building an interferometric telescope based on the single-photon multi-mode W state \cite{telescope}. The protocol has the potential to eliminate the baseline length limit, and allows in principle the interferometers with arbitrarily long baselines.

  However, in practical applications, the signals will inevitably interact with the environment during the storage and transmission process. In this way, the perfect entangled W states also may be degraded to a mixed state or a pure partially entangled states because of the environmental noise. During the applications, such partially entangled state may further decrease and cannot ultimately set up the high quality quantum entanglement channel \cite{memory}. Therefore, we need to recover the mixed state or pure partially entangled W state into the maximally entangled W state.

Here, we focus on recovering the pure partially entangled W state into the maximally entangled W state. The entanglement concentration is a powerful method to distill the maximally entangled state from the pure partially entangled state \cite{C.H.Bennett2,swapping1,swapping2,zhao1,Yamamoto1,shengpra2,shengqic,cao,zhanglihua,wanghf,gub,duff,deng4,shengPRA,zhoujosa,shengqip,zhouqip2,zhouqip1,shengpra3,dengpra,wangc1,wangc2}.  In 1996, Bennett \emph{et al.} proposed the first entanglement concentration protocol (ECP) which is known as the Schmidt projection method \cite{C.H.Bennett2}. Since then, various ECPs have been put forward successively, such as the ECP based on the entanglement swapping \cite{swapping1} and the ECP based on unitary transformation \cite{swapping2}. In 2001, Zhao \emph{ et al.} and Yamamoto \emph{et al.} proposed two similar concentration protocol independently with linear optical elements, and later realized them in experiments, respectively \cite{zhao1,Yamamoto1}. In 2008, Sheng \emph{et al.} developed their protocols with the help of the cross-Kerr nonlinearity \cite{shengpra2}.  However, most ECPs described above
are focused on the two-particles entanglement, which can not be used to concentrate the pure partially entangled W state. In 2003, Cao and Yang
firstly proposed an ECP for W state with the joint unitary transformation \cite{cao}. In 2007, Zhang \emph{et al.} proposed an ECP for the W state with the help of the collective Bell-state measurement \cite{zhanglihua}. In 2010, Wang\emph{ et al.} proposed an ECP for
a special W state as $\alpha|HVV\rangle+\beta(|VHV\rangle+|VVH\rangle)$ with linear optics \cite{wanghf}. In 2012, Gu \emph{et al.} and Du \emph{et al.} improved the ECP for the special W state with the help of the cross-Kerr nonlinearity \cite{gub,duff}. Later, Ren \emph{et al.} proposed an ECP for multipartite electron-spin states with CNOT gates \cite{deng4}. The concentration protocols for both arbitrary multi-photon partially entangled W state and
single-photon multi-mode W state were proposed \cite{shengPRA,zhoujosa,shengqip,zhouqip2}.  Unfortunately, all the previous ECPs  for partially entangled
W state are not optimal. Some of the ECPs are focused on the special types of the W states, and some ECPs need the cross-Kerr nonlinearity medium
to complete the task, which cannot be realized in current experimental conditional. Moreover, Most  ECPs cannot reach a high success probability.

 In this paper, we will  present two optimal ECPs for multi-mode single-photon W state and multi-photon polarization W state, respectively, inspired by the recent excellent concentration work for two-photon system proposed by the group of Deng \cite{dengarxiv}. Both of our two ECPs do not require any auxiliary photon, and only resort to the linear optical elements. Therefore, they can be easily realized under current experimental condition. Meanwhile, our ECPs only require local operations, which can simplify the operations largely. Moreover, our ECP only need to be operated for one time, and its success probability is higher than all the previous ECPs for W states \cite{shengPRA,zhoujosa,shengqip,zhouqip2}. Based on the features above, our ECPs may be useful in current quantum communications.

The paper is organized as follows. In Sec. 2, we first briefly explain the ECP for the  single-photon multi-mode partially entangled W state. In Sec. 3, we explain the ECP for the multi-photon polarization  partially entangled W state. In Sec. 4, we make a discussion and summary.

\section{The efficient ECP for the single-photon multi-mode W state}
\begin{figure}[!h]
\begin{center}
\includegraphics[width=10cm,angle=0]{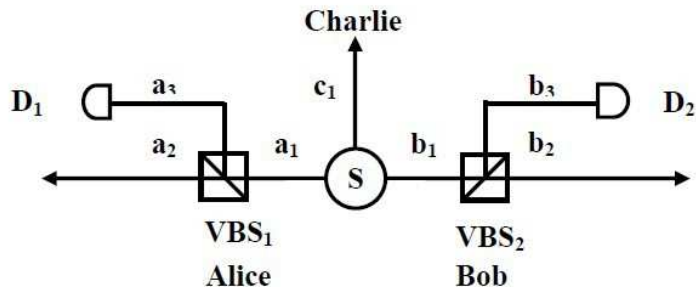}
\caption{The schematic drawing of the ECP for the single-photon multi-mode W state. The ECP can be divided into two steps. The two concentration steps are independent. Alice and Bob can operate the two steps alone, respectively. In each concentration step, a variable beam splitter (VBS) is used to adjust the entanglement coefficient.}
\end{center}
\end{figure}
Now we first start to explain our ECP for the single photon three-mode W state and then extend this method to the case of  single-photon multi-mode partially entangled W state. The basic principle of our ECP is shown in Fig. 1. Suppose a single photon source S emits a single photon, and sends it to three parties, say Alice, Bob and Charlie. In this way, it can create a single photon multi-mode W state in the spatial mode a1, b1 and c1 as
 \begin{eqnarray}
|\Phi_{1}\rangle_{a1b1c1}=\alpha|100\rangle_{a1b1c1}+\beta|010\rangle_{a1b1c1}+\gamma|001\rangle_{a1b1c1}. \label{initial}
\end{eqnarray}
Here, $\alpha,\beta$, and $\gamma$ are the initial entanglement coefficients and $|\alpha|^{2}+|\beta|^{2}+|\gamma|^{2}=1$. Meanwhile, we suppose $|\alpha|>|\beta|>|\gamma|$.

Our ECP can be divided into two steps. In the first step, Alice makes the photon in the a1 mode pass through a variable beam splitter (VBS1) with the transmittance of $t_{1}$. After VBS1, the photon state in the a1 mode evolves to
\begin{eqnarray}
|\phi\rangle_{a1}= \alpha\sqrt{t_{1}}|1\rangle_{a2}+\alpha\sqrt{1-t_{1}}|1\rangle_{a3}.
\end{eqnarray}
After passing through the VBS1, the initial state becomes
\begin{eqnarray}
|\Phi_{1}\rangle_{a1b1c1}&=&\alpha|100\rangle_{a1b1c1}+\beta|010\rangle_{a1b1c1}+\gamma|001\rangle_{a1b1c1}\nonumber\\
&\rightarrow&(\alpha\sqrt{t_{1}}|100\rangle_{a2b1c1}+\alpha\sqrt{1-t_{1}}|100\rangle_{a3b1c1})+\beta|010\rangle_{a2b1c1}+\gamma|001\rangle_{a2b1c1}.\label{initial2}
\end{eqnarray}
Then, Alice detects the photon in the a3 mode by the single photon detector D1. It is easily to found that D1 may detect one photon or no photon. Alice selects the case that D1 detects no photon. In this way, the single photon state in the three parties becomes
\begin{eqnarray}
|\Phi_{2}\rangle_{a2b1c1}=\alpha\sqrt{t_{1}}|100\rangle_{a2b1c1}+\beta|010\rangle_{a2b1c1}+\gamma|001\rangle_{a2b1c1},\label{vbs1}
\end{eqnarray}
with the success probability of $|\alpha|^{2}t_{1}+|\beta|^{2}+|\gamma|^{2}$.

It can be found that if Alice can find a suitable VBS1 with $t_{1}=\frac{|\gamma|^{2}}{|\alpha|^{2}}$, Eq. (\ref{vbs1}) can be rewritten as
\begin{eqnarray}
|\Phi_{2}\rangle_{a2b1c1}=\gamma|100\rangle_{a2b1c1}+\beta|010\rangle_{a2b1c1}+\gamma|001\rangle_{a2b1c1},\label{step1}
\end{eqnarray}
which only has two different entanglement coefficients $\gamma$ and $\beta$.

 Until now, the first concentration step is completed. In the first step, by selecting the suitable VBS with $t_{1}=\frac{|\gamma|^{2}}{|\alpha|^{2}}$ and the case that the photon detector D1 detects no photon, Alice successfully convert Eq. (\ref{initial}) to Eq. (\ref{step1}) with the success probability of
 \begin{eqnarray}
 P_{1}=2|\gamma|^{2}+|\beta|^{2},
\end{eqnarray}
where the subscript "1" means in the first concentration step.

The second concentration step is operated by Bob and the whole operation process is quite similar with the first step. Firstly, Bob makes the photon in the b1 mode pass through the VBS2 with the transmittance of $t_{2}$. After the VBS2, the photon state in the b1 mode can evolve to
\begin{eqnarray}
|\phi'\rangle_{b1}=\beta\sqrt{t_{2}}|1\rangle_{b2}+\beta\sqrt{1-t_{2}}|1\rangle_{b3}.\label{vbs2}
\end{eqnarray}
 Bob also detects the photon in the b3 mode by the single photon detector D2. When D2 detects no photon, the single photon state in the three parties can evolve to
\begin{eqnarray}
|\Phi_{3}\rangle_{a2b2c1}=\gamma|100\rangle_{a2b2c1}+\beta\sqrt{t_{2}}|010\rangle_{a2b2c1}+\gamma|001\rangle_{a2b2c1},\label{step2}
\end{eqnarray}
with the success probability of $\frac{2|\gamma|^{2}+|\beta|^{2}t_{2}}{2|\gamma|^{2}+|\beta|^{2}}$.

Similarly, if Bob can select a suitable VBS2 with $t_{2}=\frac{|\gamma|^{2}}{|\beta|^{2}}$, Eq. (\ref{step2}) can finally evolve to
\begin{eqnarray}
|\Phi\rangle_{a2b2c1}&=&\gamma|100\rangle_{a2b2c1}+\gamma|010\rangle_{a2b2c1}+\gamma|001\rangle_{a2b2c1}\nonumber\\
&\longrightarrow&\frac{1}{\sqrt{3}}(|100\rangle_{a2b2c1}+\gamma|010\rangle_{a2b2c1}+\gamma|001\rangle_{a2b2c1}),\label{max}
\end{eqnarray}
which is the maximally entangled single photon W state. When $t_{2}=\frac{|\gamma|^{2}}{|\beta|^{2}}$, the success probability of the second concentration step is
\begin{eqnarray}
P_{2}=\frac{3|\gamma|^{2}}{2|\gamma|^{2}+|\beta|^{2}},
\end{eqnarray}
where the subscript "2" means in the second concentration step.

So far, the whole ECP is completed and the three parties can finally share a maximally entangled W state from the partially entangled single photon W state. In the practical experiment, the two concentration steps are absolutely independent, which can be completed by Alice and Bob alone, respectively. The total success probability equals to the product of the success probability in each concentration step, which can be written as
\begin{eqnarray}
P_{total}=P_{1}P_{2}=(2|\gamma|^{2}+|\beta|^{2})\frac{3|\gamma|^{2}}{2|\gamma|^{2}+|\beta|^{2}}=3|\gamma|^{2}.\label{probability}
\end{eqnarray}

Similarly, it is obvious that our ECP can be extended to concentrate single photon N-mode partially entangled W state. Suppose the N-mode single photon W state is shared by N parties, which can be written as
\begin{eqnarray}
 |\Phi_{N}\rangle=a_{1}|100\cdots 0\rangle+a_{2}|010\cdots 0\rangle+a_{3}|001\cdots 0\rangle+\cdots+a_{N}|000\cdots 01\rangle,\label{initial-N}
 \end{eqnarray}
where $|a_{1}|^{2}+|a_{2}|^{2}+|a_{3}|^{2}+\cdots+|a_{N}|^{2}=1$, and $|a_{1}|>|a_{2}|>|a_{3}|>\cdots >|a_{N}|$. Under this case, N-1 parties need to perform the concentration step, respectively. In each concentration step, a suitable VBS with the transmittance of $t_{i}=\frac{|a_{n}|^{2}}{|a_{i}|^{2}}$ should be provided. After the N-1 concentration steps, Eq. (\ref{initial-N}) can be finally converted to the maximally entangled W state as
\begin{eqnarray}
 |\Phi'_{N}\rangle=\frac{1}{\sqrt{N}}(|100\cdots 0\rangle+|010\cdots 0\rangle+|001\cdots 0\rangle+\cdots+|000\cdots 01\rangle),\label{max-N}
 \end{eqnarray}
with the success probability of
\begin{eqnarray}
 P_{N_{total}}=N|a_{N}|^{2}.\label{probability-N}
 \end{eqnarray}

\section{The ECP for the multi-photon polarization W state}
\begin{figure}[!h]
\begin{center}
\includegraphics[width=10cm,angle=0]{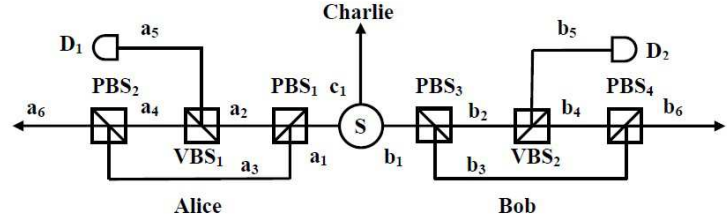}
\caption{The schematic drawing of the ECP for the three-photon polarization W state. The ECP also can be divided into two independent steps, which only requires local operations from Alice and Bob, respectively. In each step, the polarization beam splitters (PBSs) are used to transmit the $|H\rangle$ polarization photon state and reflect the $|V\rangle$ polarization photon state. The VBSs are used to adjust the entanglement coefficients.}
\end{center}
\end{figure}
Interestingly, with the basic principle in Sec. 2, we can still propose an efficient ECP for concentrating the partially entangled multi-photon polarization state. We take the three-photon W state as an example. The basic principle of the ECP is shown in Fig. 2. Suppose a single photon source S emits three photons and sends them to Alice, Bob and Charlie, respectively, which creates a partially entangled three-photon W state in a1, b1 and c1 modes as
\begin{eqnarray}
|\Psi_{1}\rangle_{a1b1c1}=\alpha|HVV\rangle_{a1b1c1}+\beta|VHV\rangle_{a1b1c1}+\gamma|VVH\rangle_{a1b1c1}. \label{initial2}
\end{eqnarray}
Here, $|\alpha|^{2}+|\beta|^{2}+|\gamma|^{2}=1$ and we also suppose $|\alpha|>|\beta|>|\gamma|$.

The ECP also can be divided into two steps. In the first step, Alice firstly maks the photon in the a1 mode pass through a polarization beam splitter (PBS), here named PBS1, which can transfer a $|H\rangle$ polarization photon and reflet a $|V\rangle$ polarization photon. After PBS1, $|\Psi_{1}\rangle_{a1b1c1}$ can evolve to
\begin{eqnarray}
|\Psi_{2}\rangle=\alpha|HVV\rangle_{a2b1c1}+\beta|VHV\rangle_{a3b1c1}+\gamma|VVH\rangle_{a3b1c1}.\label{PBS1}
\end{eqnarray}

Then Alice makes the photon in the a2 mode pass through a variable beam splitter (VBS1) with the transmittance of $t_{1}$. In this way, Eq. (\ref{PBS1}) can evolve to
\begin{eqnarray}
|\Psi_{2}\rangle=\alpha\sqrt{t_{1}}|HVV\rangle_{a4b1c1}+\alpha\sqrt{1-t_{1}}|HVV\rangle_{a5b1c1}+
\beta|VHV\rangle_{a3b1c1}+\gamma|VVH\rangle_{a3b1c1}.\label{PBS1-1}
\end{eqnarray}
After that, Alice detects the photon in the a5 mode by the single photon detector D1. If D1 detects no photon, Eq. (\ref{PBS1-1}) will collapse to
\begin{eqnarray}
|\Psi_{2}\rangle=\alpha\sqrt{t_{1}}|HVV\rangle_{a4b1c1}+\beta|VHV\rangle_{a3b1c1}+\gamma|VVH\rangle_{a3b1c1},\label{PBS1-2}
\end{eqnarray}
with the possibility of $|\alpha|^{2}t_{1}+|\beta|^{2}+|\gamma|^{2}$. Similar with Sec. 2, if $t_{1}=\frac{|\gamma|^{2}}{|\alpha|^{2}}$, Eq. (\ref{PBS1-2}) can be written as
\begin{eqnarray}
|\Psi_{2}\rangle\rightarrow\gamma|HVV\rangle_{a4b1c1}+\beta|VHV\rangle_{a3b1c1}+\gamma|VVH\rangle_{a3b1c1},\label{PBS1-3}
\end{eqnarray}
which only has two different entanglement coefficients $\beta$ and $\gamma$.

Finally, Alice makes the photon in the a3 and a4 mode pass through another PBS, here named PBS2. After PBS2, Eq. (\ref{PBS1-3}) evolves to
\begin{eqnarray}
|\Psi_{3}\rangle_{a6b1c1}=\gamma|HVV\rangle_{a6b1c1}+\beta|VHV\rangle_{a6b1c1}+\gamma|VVH\rangle_{a6b1c1}. \label{2-step1}
\end{eqnarray}

Until now, the first concentration step is completed and we successfully obtain the three-photon W state with only two different entanglement coefficients, with the success probability of $P_{1}=|\beta|^{2}+2|\gamma|^{2}$.

The second concentration step is operated by Bob alone, which is quite similar with the first concentration step. As shown in Fig. 2, by making the photon in the b1 mode pass through PBS3 and the photon in b2 mode pass through the VBS2 with the transmittance of $t_{2}$, Eq. (\ref{2-step1}) can ultimately evolve to
\begin{eqnarray}
|\Psi_{4}\rangle=\gamma|HVV\rangle_{a6b3c1}+\beta\sqrt{t_{2}}|VHV\rangle_{a6b4c1}+\beta\sqrt{1-t_{2}}|VHV\rangle_{a6b5c1}
+\gamma|VVH\rangle_{a6b3c1}.\label{2-PBS1}
\end{eqnarray}
Then, the photon in the b5 mode is detected by the single photon detector D2. Under the case that D2 detects no photon, Eq. (\ref{2-PBS1}) will collapse to
\begin{eqnarray}
|\Psi_{4}\rangle=\gamma|HVV\rangle_{a6b3c1}+\beta\sqrt{t_{2}}|VHV\rangle_{a6b4c1}+\gamma|VVH\rangle_{a6b3c1},\label{2-PBS2}
\end{eqnarray}
with the probability of $t_{2}$.
If $t_{2}=\frac{|\gamma|^{2}}{|\beta|^{2}}$, Eq. (\ref{2-PBS2}) can finally be written as
\begin{eqnarray}
|\Psi_{5}\rangle=\frac{1}{\sqrt{3}}(|HVV\rangle_{a6b3c1}+|VHV\rangle_{a6b4c1}+|VVH\rangle_{a6b3c1}).
\end{eqnarray}
Finally, Bob makes the photon in the b3 and b4 modes pass through the PBS4. After the PBS4, the three parties can share a maximally entangled polarization W state as
\begin{eqnarray}
|\Psi_{6}\rangle=\frac{1}{\sqrt{3}}(|HVV\rangle_{a6b6c1}+|VHV\rangle_{a6b6c1}+|VVH\rangle_{a6b6c1}).
\end{eqnarray}

The total success probability of the ECP also equals the product of the success probability in each concentration round, which is the same as that in Eq. (\ref{probability})

Similarly, by performing N-1 concentration steps described above, our ECP can also be extended to concentrate the partially entangled N-photon polarization W state as
 \begin{eqnarray}
 |\Psi_{N}\rangle=a_{1}|HVV\cdots V\rangle+a_{2}|VHV\cdots V\rangle+a_{3}|VVH\cdots V\rangle+\cdots+a_{N}|VVV\cdots VH\rangle,\label{initial-N2}
 \end{eqnarray}
where $|a_{1}|^{2}+|a_{2}|^{2}+|a_{3}|^{2}+\cdots +|a_{4}|^{2}=1$, and $|a_{1}|>|a_{2}|>|a_{3}|>\cdots >|a_{N}|$. With the help of the PBSs and suitable VBS in each concentration step, Eq. (\ref{initial-N2}) can be finally recovered to the maximally entangled N-photon polarization W state as
\begin{eqnarray}
|\Psi'_{N}\rangle=\frac{1}{\sqrt{N}}(|HVV\cdots V\rangle+|VHV\cdots V\rangle+|VVH\cdots V\rangle+\cdots+|VVV\cdots VH\rangle),\label{initial-N2}
 \end{eqnarray}
 with the same success probability in Eq. (\ref{probability-N}).

\section{Discussion and summary}
\begin{figure}[!h]
\begin{center}
\includegraphics[width=8cm,angle=0]{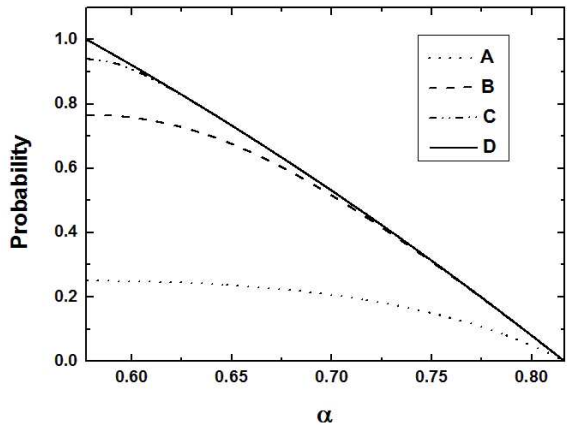}
\caption{The total probability P for obtaining a maximally entangled three-photon polarization W state of the ECPs in Ref \cite{shengPRA,zhoujosa,shengqip,zhouqip2} (curve A, B, C) and our paper (curve D), which is altered with the initial coefficient $\alpha$. Here, we choose $\beta=\frac{1}{\sqrt{3}}$. In the current ECP, we suppose $|\alpha|>|\beta|>|\gamma|$, so that we make $\alpha\in(\sqrt{\frac{1}{3}},\sqrt{\frac{2}{3}})$. As the ECPs in Ref \cite{shengPRA,zhoujosa,shengqip,zhouqip2} can be used repeatedly to further concentrate the partially entangled W state, we make curve A represents both the two steps are operated for one time, curve B represents both two steps are operated for three times, and curve C represents both the two steps are operated for five times. Curve D represents the ECP in our current paper.}
\end{center}
\end{figure}
In the paper, we propose two efficient ECPs for partially entangled multi-mode single photon W state and multi-photon polarization W state. Our ECPs only require the linear optical elements, among which the VBS is the key elements. We require the VBSs with suitable transmittance to adjust the entanglement coefficients and finally obtain the maximally entangled W state. Actually, the VBS is a common linear optical element in current experiment conditions. Recently, Osorio \emph{et al.} reported their results about the photon amplification for quantum communication with the help of the VBSs \cite{amplification}.
 In their paper, they increased the probability $\eta_{t}$ of the single photon $|1\rangle$ from a mixed state as $\eta_{t}|1\rangle\langle1|+(1-\eta_{t})|0\rangle\langle0|$ with the help of the VBSs. In their experiment, they successfully adjust the transmittance of the VBSs from 50:50 to 90:10 to increase the visibility from 46.7 $\pm$ 3.1\% to 96.3 $\pm$ 3.8\%. This result ensures our ECP can be realized under current experimental conditions. Meanwhile, in our ECPs, each concentration step only requires local operation, which can simplify the experimental operation largely. On the other hand, in linear optics, when the photon is detected by the detector, it will be destroyed, which is well known as the post selection principle. In our ECPs, as each party only selects the case that the photon detector measures no photon, the generated maximally entangled W state will not be destroyed, and can be used in other applications.

Moreover, although our current ECPs can not be recycled, their success probability is higher than the previous ECPs for the W state \cite{shengPRA,zhoujosa,shengqip,zhouqip2}. Now, we will compare the success probability of our current ECPs with our previous ECPs for W state. We just take the three-photon polarization W state in Eq. (\ref{initial2}) as an example. In our previous papers \cite{shengPRA}, the ECPs also contain two concentration steps. In each step, we require an auxiliary single photon. Meanwhile, the ECPs can be used repeatedly to further concentrate the partially entangled W state. The success probability of the two concentration steps can be written as
  \begin{eqnarray}
P_{N}^{1}&=&\frac{|\alpha|^{2^{N}}(|\beta|^{2^{N}-2}|\gamma|^{2}+2|\beta|^{2^{N}})}{(|\alpha|^{2^{N}}+|\beta|^{2^{N}})(|\alpha|^{2^{N-1}}
+|\beta|^{2^{N-1}})\cdots(|\alpha|^{2}+|\beta|^{2})},\nonumber\\
P_{M}^{2}&=&\frac{3\beta^{2^{M}}\gamma^{2^{M}}}{(\gamma^{2^{M}}+\beta^{2^{M}})(\gamma^{2^{M-1}}+\beta^{2^{M-1}})\cdots(\gamma^{2}+\beta^{2})}
\cdot\frac{1}{(\gamma^{2}+2\beta^{2})},
\end{eqnarray}
where the superscript "1" and "2" mean in the first and second concentration step, respectively. The subscripts "N" and "M" mean in the Nth and Mth concentration round.

Therefore, by repeating both steps, the total success probability is
\begin{eqnarray}
P_{total}&=&P_{1}^{1}(P_{1}^{2}+P_{2}^{2}+\cdots+P_{M}^{2}
+P_{2}^{1}(P_{1}^{2}+P_{2}^{2}+\cdots+P_{M}^{2})\nonumber\\
&+&\cdots +P_{N}^{1}(P_{1}^{2}+P_{2}^{2}+\cdots+P_{M}^{2})\nonumber\\
&=&\sum_{N=1}^{\infty}P^{1}_{N}\sum_{M=1}^{\infty}P^{2}_{M}.\label{total}
\end{eqnarray}

Here, we calculate the total success probability of both our current ECPs in Eq. (\ref{probability}) and the previous ECP in Eq. (\ref{total}) in Fig. 3. Here, we choose $\beta=\frac{1}{\sqrt{3}}$. In the current paper, we suppose $|\alpha|>|\beta|>|\gamma|$, so that we make $\alpha\in(\sqrt{\frac{1}{3}},\sqrt{\frac{2}{3}})$. In Fig. 3, curves A, B, and C represent the total success probability of the ECP in Ref. \cite{shengPRA}. Curve A represents that both the two steps are operated for one time. Curve B represents that both two steps are operated for three times. Curve C represents that both the two steps are operated for five times. Curve D represents the ECP in the current paper. It can be found that in both two ECPs, the success probability is largely altered with the initial entanglement coefficient $\alpha$. The higher initial entanglement can obtain the higher success probability. Moreover, although the success probability of the ECP in Ref. \cite{shengPRA} increases with the cycle times, it is still lower than that of our current ECPs. Especially, when $\alpha=\frac{1}{\sqrt{3}}$, the success probability of our current ECP is 1, while that of the ECP in Ref. \cite{shengPRA} can only obtain about 0.93, when both two concentration steps are operated for five times. Certainly, we can further increase its success probability by increasing its cycle times. However, by mathematical calculation, we can get when the ECP in Ref. \cite{shengPRA} is repeated indefinitely, its success probability curve (which will not be presented in Fig. 3) will be coincided with curve D. During our ECPs, the total success probability essentially is decided by the smallest coefficient of the initial state.

Actually, in the early theoretical work of concentration of the two-particle Bell state, Lo
and Popescu showed that the maximum probability
with which a Bell state can be obtained by purifying a single
entangled pair is twice the modulus square
of the Schmidt coefficient of smaller magnitude \cite{lo}. The result of the recent work of Deng's group is consist with  Lo
and Popescu \cite{dengarxiv}. It reveals that the total entanglement is a  conserved quantity. Interestingly, our ECPs can be regarded as the
extension of the result from the previous work of two-particle case, which can be concluded as the maximum probability of concentrating a N-particle partially entangled W
state or single-photon N-mode partially entangled state is Nth the modulus square of the Schmidt coefficient of the smallest magnitude.

In summary, we proposed two optimal ECPs for concentrating the single-photon multi-mode W state and N-photon polarization W state. In both ECPs, we only require one pair of partially entangled W state, and do not consume any auxiliary photon. In each concentration step, we mainly require the VBS to adjust the entanglement coefficients. Our ECPs have some obvious advantages. First, they only require the linear optical elements, which makes them can be easily realized under current experimental condition. Second, the generated maximally entangled W state will not be destroyed, and can be used in other applications. Third, our ECPs only need to be operated for one time, but they can obtain higher success probability than previous ECPs. Based on the advantages above, our ECPs may be useful in current quantum communication fields.

\section{Acknowledgements}
 The project is supported by the National Natural Science Foundation of China
(Grant No. 11104159), the Open Research Fund Program of National
 Laboratory of Solid State Microstructures, Nanjing University (Grant No. M25022), the Open Research Fund Program of the State Key Laboratory of
Low-Dimensional Quantum Physics, Tsinghua University (Grant No. 20120904), and the open research fund of Key Lab of Broadband Wireless Communication and Sensor Network Technology, Nanjing University of Posts and Telecommunications, Ministry of Education (No. NYKL201303), and the Priority Academic Program Development of Jiangsu Higher Education Institutions.

\end{document}